\documentclass[12pt]{iopart}
\usepackage{psfrag}
\usepackage{graphicx}% Include figure files
\usepackage{dcolumn}% Align table columns on decimal point
\usepackage{bm}% bold math
\usepackage{epsfig}
\usepackage{yfonts}
\usepackage{version}
\usepackage[isolatin]{inputenc}
\newcommand{\bra}[1]{\langle\,{#1}\, |}
\newcommand{\ket}[1]{|\,{#1}\,\rangle}
\newcommand{\braket}[2]{\mbox{$\langle\,{#1}\, | \,{#2}\,\rangle$}}

\newcommand{\HamTot}{H}
\newcommand{\HamEl}{H^{\rm el}}
\newcommand{\Vdd}{V}

\newcommand{\eTrans}{\varepsilon}
\newcommand{\adEnergy}{U}
\newcommand{\adBasis}{\phi}

\newcommand{\Ort}{R}

\newcommand{\Abstand}{x}
\newcommand{\altAbstand}{y}
\newcommand{\Pos}{\Ort}
\newcommand{\PosAbs}{\Ort}

\newcommand{\Zustand}{\psi}

\newcommand{\Kraft}{F}

\newcommand{\Ekin}{E_{\rm kin}}
\newcommand{\Konfig}{\vec{\rho}}
\newcommand{\vecd}{\vec{d}}

\excludeversion{weg}

\begin{document}

\title{Motion of Rydberg atoms induced by resonant dipole-dipole interactions}% Force line breaks with \\
 \author{C Ates$^1$, A Eisfeld$^2$ and J M Rost$^1$}
\address{$^1$ Max Planck Institute for the Physics of Complex Systems
N\"othnitzer Stra\ss e 38, 01187 Dresden, Germany }
\address{$^2$ Theoretical Quantum Dynamics, Universit\"at Freiburg,
Hermann-Herder-Str. 3,  79104 Freiburg, Germany} 
\ead{cenap@pks.mpg.de}

%\affiliation{%
%Theoretische Quantendynamik, Universit\"at Freiburg,
%Hermann-Herder-Str. 3, D-79104 Freiburg, Germany
%}%

\date{\today}% It is always \today, today,

\begin{abstract}
We show that nuclear motion of Rydberg atoms can be induced by resonant
dipole-dipole interactions that trigger the energy transfer between two
energetically close Rydberg states.  How and if the atoms move depends on
their initial arrangement as well as on the initial electronic excitation.
Using a mixed quantum/classical propagation scheme we obtain the
trajectories and kinetic energies of atoms, initially arranged in a regular
chain and prepared in excitonic eigenstates.  The influence of off-diagonal
disorder on the motion of the atoms is examined and it
is shown that  irregularity in the arrangement of the atoms can lead to
an acceleration of the nuclear dynamics.
\end{abstract}

\maketitle

\section{Introduction}
Resonant energy transfer due to dipole-dipole interaction  plays a crucial role in many areas of physics, chemistry, material sciences and biology. Prominent examples are the light harvesting units of plants and bacteria \cite{AmVaGr00__}, aggregates consisting of organic dyes \cite{Ko96__,WeSt03_125201_} or molecular crystals \cite{Da62__}.
Since the work of Franck and Teller \cite{FrTe38_861_} the theoretical framework to treat such processes is based on Frenkel's exciton concept \cite{Fr31_1276_}.
However, internal and external degrees of freedom, e.g., internal vibrational modes of molecules and interactions with the surrounding solvent or proteins, have great influence on the properties of those systems and render a complete theoretical description challenging \cite{EiBr06_113003_}.
Hence, a prototypical system for a systematic study of the effect of dipole-dipole interaction should have  (i) a simple structure,  (ii) should not couple strongly to the environment and (ii) should be easy to  manipulate, e.g. in terms of its interaction strength or geometrical arrangement.

Rydberg atoms in ultracold gases are promising candidates to meet these requirements:  First of all, their internal structure can often be reduced to a few relevant energy levels. Second, they posess long radiative lifetimes and  their coupling to the environment via thermal collisions plays a minor role due to the low thermal velocities in ultracold gases. Finally, their mutual interactions can be controlled precisely, since the transition dipole moment of two adjacent Rydberg states increases quadratically with the principal quantum number.

Consequently, clear evidences for resonant energy transfer due to dipole-dipole interaction have been observed in ultracold (''frozen'') Rydberg gases 
\cite{MoCoTo98_253_,AnVeGa98_249_,AfBoVa04_233001_,WeAmOl06_37_}.
There have been also several proposals how to study energy transfer in regular arrangements of Rydberg atoms \cite{RoHeTo04_042703_,MueBlAm07_090601_}.
These geometries could in principle be produced by using optical lattices, or microlens arrays. 
Such one-dimensional chains have served as a fruitfull model to understand spectral and energy transfer properties of dipole-dipole interacting systems and to investigate the influence of disorder on  their electronic transport properties.

In studies on energy transfer the atoms/molecules are usually considered to remain at their initial positions.
However, similar to the van der Waals interaction \cite{AmReWe07_023004_,AmReGi07_054702_}, the dipole-dipole interaction can lead to forces between Rydberg atoms, accelerating the particles. This effect has to be taken into account because it can lead to ionizing collisions and subsequent plasma formation, as demonstrated experimentally 
\cite{LiTaGa05_173001_,LiTaJa06_27_,FiCoDr99_1839_}. 
However, the induced forces might also be used to control the motion of the atoms, as recently  suggested \cite{WaRoJo07_3693_}.

In this work, we will investigate the motion induced  by the resonant dipole-dipole interaction that triggers the  energy transfer processes between atoms in energetically close Rydberg states, as observed in the studies on ''frozen'' Rydberg gases.
To this end, we will assume that the Rydberg atoms are arranged in a linear array and focus on two distinct  situations. 
First, we will consider a regular chain, in which all the Rydberg atoms are prepared such that they initially have equal nearest neighbor distance and second, the situation when the Rydberg atom at one of the ends of the chain has a smaller distance to its neighbor than the atoms in the rest of the chain. 
The latter situation is a simple example, where the regularity of the mutual interactions in the system is broken.  Such a system is often refered to as exhibiting ''off-diagonal disorder''.

In both cases we will assume, that the system is initially prepared in one of the electronic eigenstates that form the exciton band.
Experimentally, this could be realized by exciting atoms arranged in a microtrap array to a specific Rydberg state and subsequently promoting this system into an excitonic eigenstate of the chain by microwave excitation, similar to the procedure used in Refs.~\cite{LiTaGa05_173001_,LiTaJa06_27_}.

A fully quantum mechanical treatment of this problem for long chains (i.e., many atoms) is computationally demanding. Therefore, we will use a mixed quantum/classical surface hopping method, in which the nuclear motion is treated classically while the electronic dynamics is treated quantum mechanically. 

The paper is organized as follows:
In section \ref{model} we give a brief description of the model system and the propagation scheme which we use.
For the regular chain, we discuss analytical estimates for the initial forces as well as for the kinetic energy of the system  and present numerical results in section \ref{regular}.
Having analyzed the basic properties of the atomic motion in the regular case, we proceed in section \ref{disorder} with the case of broken regularity, by placing two atoms at one end of the chain closer together. Finally, section \ref{concl} presents our conclusions.

Throughout this work atomic units will be used, unless stated otherwise.
%%%%%%%%%%%%%%%%%%%%%%%%%%%%%%%%%%%%%%%%%%%%%%%%%%%%%%%%%%%%%%%%%%%%%%%%%%%%%%%%%%%%%%%%%%%%%%%%%%%%%%%%%%%%%%%%%%%%%%%%
\section{Rydberg atoms on a one dimensional chain}
\label{model}
%%%%%%%%%%%%%%%%%%%%%%%%%%%%%%%%%%%%%%%%%%%%%%%%%%%%%%%%%%%%
\subsection{Formulation of the model}
We consider a chain of $N$ identical atoms with mass $M$ and denote the position of the $n$th atom by $\Pos_n$.
The  distance $\PosAbs_{nm}\equiv | \Pos_m-\Pos_n|$  between the atoms is assumed to be large enough that an 
overlap of their nuclear and  electronic wavefunctions can safely be neglected. 
Since the focus of the present work is on resonant transitions between two energetically close Rydberg states, an essential state picture will be used, where for every atom only two electronic states are considered.
We denote the energetically lower lying (Rydberg) state by $|s \rangle$ and the upper one by $|p \rangle$ and sometimes refer to the latter as the ''excited state'' or the ''excitation''. The energies of these states are $\eTrans_s$ and  $\eTrans_p$ respectively, and $\mu$ denotes the transition dipole between $|s \rangle$ and $|p \rangle$. 
By
\begin{equation}
\label{def:pi_n}
\ket{\pi_n}\equiv \ket{s\cdots p\cdots s}
\end{equation}
we denote a state in which atom $n$ is in state $\ket{p}$ and all others are in state $\ket{s}$.
Without any interaction between the atoms the $N$ states defined in \eref{def:pi_n} are degenerate eigenstates of the electronic system   with energy  $\eTrans=\eTrans_p+(N-1)\eTrans_s$ which we choose to be the zero of energy and restrict our  discussion to the space spanned by the states \eref{def:pi_n} in the following.

The electronic energy transfer between two atoms $n$ and $m$ at distance $\PosAbs_{nm}$ 
 (Fig \ref{fig:theory}(a)) is mediated by the dipole-dipole interaction
\begin{equation}
\label{vdd}
\Vdd_{nm}(\Pos_{nm}) = \frac{\mu^2}{\PosAbs_{nm}^3} \, ,
\end{equation}
where for the sake of simplicity we have ignored orientational effects.
Due to this resonant dipole-dipole interaction the states given by (\ref{def:pi_n}) are no eigenstates of the electronic system anymore.
Moreover, as the coupling (\ref{vdd}) depends explicitly on the positions of the atoms, their center of mass motion will also be affected by the dipole-dipole interaction, i.e., a force will be exerted on the particles that is proportional to the gradient of $\Vdd_{nm}$ with respect to the nuclear coordinates of the atoms.
Consequently, both subsystems (electronic and center of mass) are coupled by (\ref{vdd}) so that  the motion of the particles will depend on the electronic state of the system and vice versa.

The total Hamiltonian for a one dimensional chain of $N$ Rydberg atoms is, thus,  given by
\begin{equation}
\label{Ham_Tot}
\HamTot(\Pos_1,\dots,\Pos_N)=-\sum_{n=1}^N\frac{\hbar^2}{2 M} \nabla^2_{\Pos_n} + H^{\rm el}(\Pos_1,\dots,\Pos_N ) \, ,
\end{equation}
where the electronic Hamiltonian, which depends on the  nuclear coordinates, is
\begin{equation}
\label{H_el}
\HamEl(\Pos_1,\dots,\Pos_N) = - \sum_{n,m =1}^N\Vdd_{nm}(\PosAbs_{nm})\ket{\pi_n}\bra{\pi_m} \, ,
\end{equation}
and we set $\Vdd_{nn} \equiv \epsilon =0$.
%%%%%%%%%%%%%%%%%%%%%%%%%%%%%%%%%%%%%%%%%%%%%%%%%%%%%%%%%%%%
\subsection{Molecular dynamics with quantum transitions}
%------------------------------------------------------------------------------------------------------------------------------------------
\begin{figure}
\centering
\includegraphics[width=0.25\textwidth]{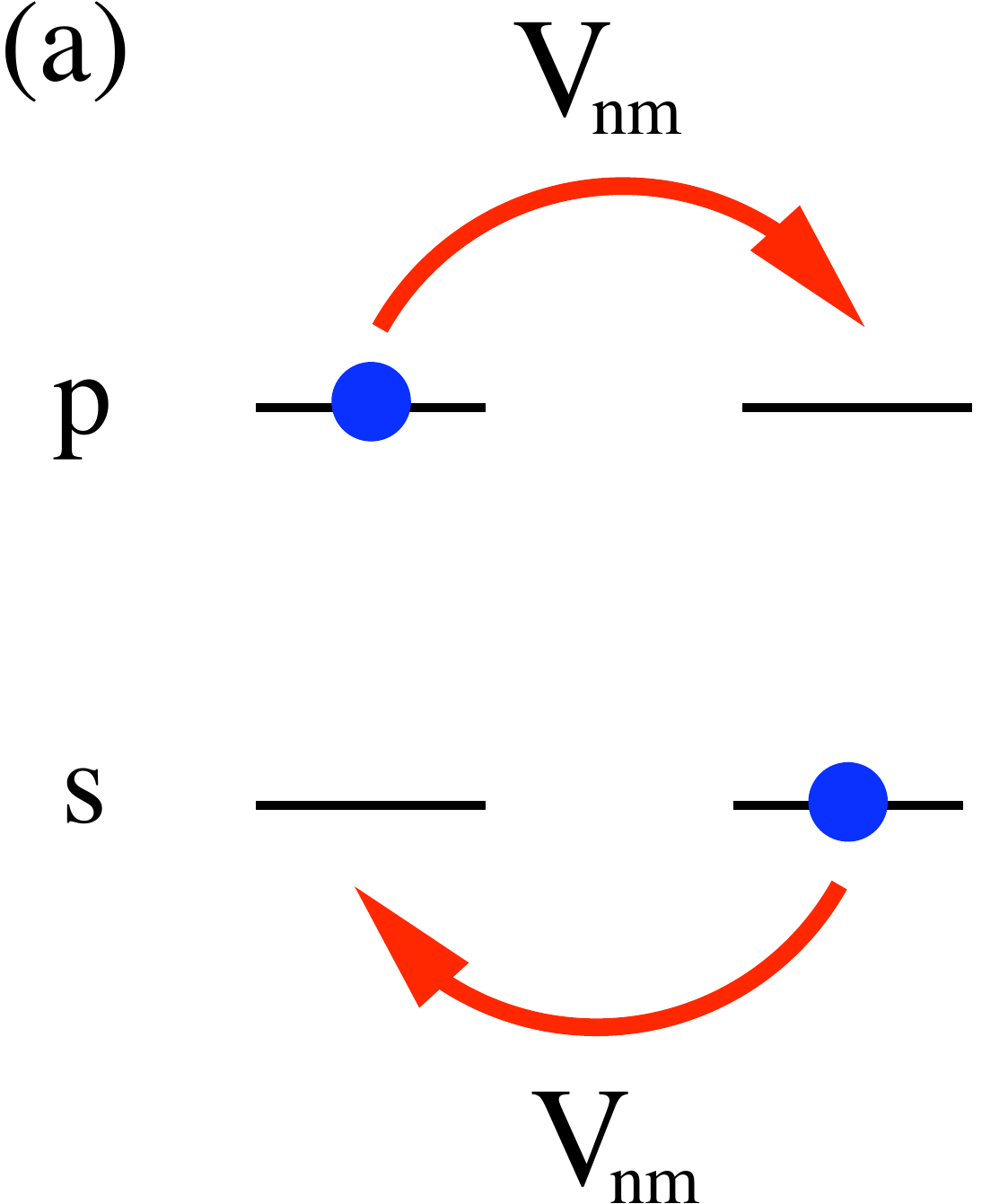} \hfill
\includegraphics[width=0.55\textwidth]{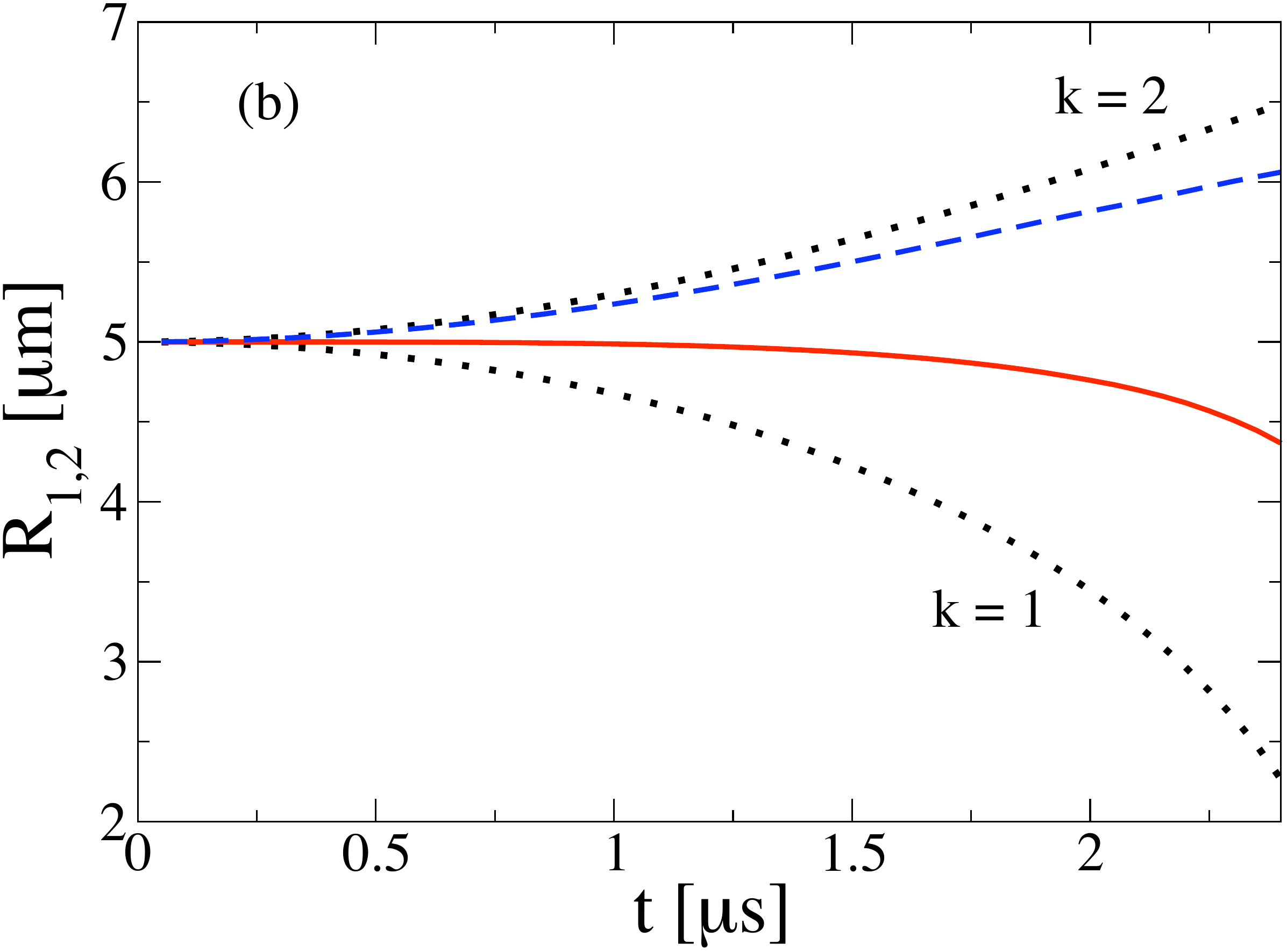}
\caption{
(a) The $p$ state is transfered from atom to atom, due to the dipole-dipole interaction $V_{nm}$.
(b) Time evolution of the atomic distance in a dimer for different initial electronic configurations. The dotted lines are the trajectories, when the particles are propagated on one of the adiabatic potentials  $U_k$ with $k=1$ and $k=2$, respectively. The solid line is the trajectory, when the electronic excitation is initially completely localized on one of the atoms, the dashed curve, when 80 per cent of the electronic excitation is localized on one of the particles and both $c_k$ (cf. Eq.\ref{ad_expansion}) have the same phase.  Parameters are $\mu = 1000$ a.u., $M=11000$ a.u. (i.e. Lithium).
\label{fig:theory}
}
\end{figure}
%------------------------------------------------------------------------------------------------------------------------------------------
 To study the motion of the Rydberg atoms induced by the dipole-dipole coupling we use  a mixed quantum/classical method, where the nuclear motion is treated classically. 
To this end we denote the configuration of all atomic positions at a given time by $\Konfig \equiv \{ \Ort_1, \dots , \Ort_N \}$ and define adiabatic electronic states $\adBasis_k(\Konfig)$ and energies $\adEnergy_k(\Konfig)$ as the eigenstates and the corresponding eigenenergies of the electronic Hamiltoinan for a given $\Konfig$,
 \begin{equation}
\label{ad_evek}
H^{\rm el}(\Konfig) \, \adBasis_k(\Konfig)=\adEnergy_k(\Konfig) \, \adBasis_k(\Konfig) \, .
\end{equation}
Expanding the total wavefunction of the system in the adiabatic basis defined above,
\begin{equation}
\label{ad_expansion}
\Zustand(\Konfig,t)=\sum_{k=1}^N c_k(t) \adBasis_k(\Konfig) \, ,
\end{equation}
and substituting (\ref{ad_expansion}) into the time-dependent Schr\"odinger
equation of the full system we arrive at $N$ coupled evolution equations for the quantum amplitudes $c_k$,
\begin{equation}
\label{class_qm_general}
i \hbar \dot{c}_k= \adEnergy_{k} c_k -i \hbar \sum_{j=1}^N  \dot{\Konfig}\cdot \vecd_{kj}\, c_j \, .
\end{equation}
The terms that couple these equations are proportional to the velocities of the particles and to the so called ''non-adiabatic coupling vector''  $\vecd_{kj}  \equiv\braket{\adBasis_k}{\nabla_{\Konfig}\adBasis_j}$ which, using the Hellmann-Feynman theorem, can be written as
\begin{equation}
 \vecd_{kj}=
\frac{\bra{\adBasis_k}\nabla_{\Konfig}H^{\rm el}(\Konfig)\ket{\adBasis_j}}{\adEnergy_j(\Konfig)-\adEnergy_k(\Konfig)} \, ,
\end{equation}
for $j\ne k$ and $\vecd_{kk}= 0$. 

For a chain of length $N=2$ (''dimer'') all non-adiabatic coupling vectors vanish identically. In this case, the classical trajectories can be obtained by propagating the atoms on each of the two adiabatic potentials $\adEnergy_k$ according to the classical equations of motion
\begin{equation}
\label{eq:motion}
M \ddot{\Pos}_n = \Kraft_n=-\nabla_{ R_n}\bra{\adBasis_k}H^{\rm el}\ket{\adBasis_k} \, ,
\end{equation}
and averaging the results with respect to the initial probabilities $|c_k(t=0)|^2$ to be in one of the adiabatic states $k$. 
Fig.\ref{fig:theory}(b) shows the trajectories for a dimer, when the motion of the atoms starts on the adiabatic potential curves and two examples, in which the system is initially prepared in different superpositions of the electronic eiegnstates. 
From these simple examples it is already seen that the atomic motion induced by the dipole-dipole interaction depends crucially on the initial electronic state of the system.

For chains with $N \ge 3$ the non-adiabatic couplings are generally nonzero.
 In this case we use a propagation scheme, in which an ensemble of trajectories is propagated, and each trajectory moves classically on a single adiabatic surface according to Eq.(\ref{eq:motion}) except for the possibility of instantaneous switches among the adiabatic states  $\ket{\adBasis_k}$ due to the presence of the $\vecd_{kj}$. 
The algorithm used in this work to correctly apportion trajectories among the states according to their quantum probabilities with the minimum required number of quantum transitions  is Tully's fewest switching algorithm \cite{Tu90_1061_}.
For initial conditions that lead to fast collisions of atoms, the non-adiabatic couplings become significant, as they are proportitional to the velocities of the particles (c.f. Eq. \ref{class_qm_general}). This is manifest in large deviations of single trajectories from the ensemble average. Therefore, the mean trajectories for such initial conditions were determined up to the point $t_{\rm coll}$, at which the fastest collision event in a single realization occured.

The numerical results that we will present in the following, have been obtained with the parameters $M=11000\;$a.u. (i.e., Lithium), $\mu=1000\;$a.u. (corresponding to typical transition dipole moments between energetically close Rydberg states with principal quantum numbers $n \approx 30, \dots, 40$) and an initial nearest neighbor distance of $\Abstand = 5\, \mu$m.

%%%%%%%%%%%%%%%%%%%%%%%%%%%%%%%%%%%%%%%%%%%%%%%%%%%
%%%%%%%%%%%%%%%%%%%%%%%%%%%%%%%%%%%%%%%%%%%%%%%%%%%
\section{Rydberg atoms arranged on a regular chain}
\label{regular}
First, we will consider a chain consisting of $N$ atoms, initially at rest with
equal spacing $\Ort_{n,n+1}\equiv \Abstand$, and we will assume that the initial electronic state of the system is an eigenstate of $\HamEl$.
If one restricts the interaction of each particle to its nearest neighbors, it
is possible to obtain a simple analytical estimate of the force (\ref{eq:motion}) acting initially
on a particular atom $n$,
\begin{equation}
\label{F_general}
\Kraft_n(t=0) \approx 2 \frac{\partial}{\partial \Abstand} V(\Abstand) \;
\bra{\adBasis_k}\pi_n\rangle \, \Big[ \bra{\adBasis_k} \pi_{n+1}\rangle -
\bra{\adBasis_k} \pi_{n-1}\rangle \Big] \, ,
\end{equation}
where we set $\bra{\adBasis_k}\pi_0\rangle = \bra{\adBasis_k}\pi_{N+1}\rangle =
0$ to account for the atoms at the ends of the chain. Thus, the initial force on particle $n$ depends on the value of the electronic eigenstate $\bra{\adBasis_k} \pi_{n\pm1}\rangle$ at both neighboring sites.

For a regular chain with nearest neighbor interactions the eigenenergies $\adEnergy_k$ and eigenstates $\ket{\adBasis_k}$ are given by (see e.g.\ Ref.~\cite{AmVaGr00__}~chapter~6)
\begin{eqnarray}
\adEnergy_k&= -2 V(\Abstand) \cos\left(\frac{\pi k}{N+1}\right) \, ,\\
\ket{\adBasis_k}&=\sqrt{\frac{2}{N+1}}\sum_{n=1}^N
\sin\left(\frac{\pi k n}{N+1}\right)\ket{\pi_n} \, .
\label{eigen_vek}
\end{eqnarray}
The eigenenergies form a band, where for $N\rightarrow\infty$ the ground state
($k=1$) has energy $-2 V(\Abstand)$ and the highest eigenstate ($k=N$) has energy $2V(\Abstand)$.
If the system is prepared in an electronic eigenstate $\ket{\adBasis_k}$,
the force  on atom $n$ can be obtained using (\ref{F_general}), 
\begin{equation}
\label{F_chain}
\Kraft_n(t=0)= \frac{12}{N+1}\, 
\frac{\mu^2}{\Abstand^4}\, 
\sin{\left(\frac{\pi k}{N+1}\right)}\,
\sin\left(\frac{2 \pi k n}{N+1}\right) \, .
\end{equation}
Hence, the force that initially acts on an atom decreases rapidly with
increasing chain length $N$.
This decrease of the magnitude is reasonable, since in the eigenstates
\eref{eigen_vek} the excitation is delocalized more or less over the whole
chain. Therefore, the excitation is shared among all atoms in the chain, i.e., a
particular atom carries only a fraction of approximately $1/N$ of the electronic
excitation.

The directions of the initial forces follow a simple scheme, that can be deduced using the symmetries of the system reflected in the nodal structure of the initial electronic wavefunction. 
Generally, the motion of the particles  will be symmetric arround the center of mass coordinate of the chain (which is a constant of motion).
In particular, for a chain with an odd number of atoms $N$ the central atom ($n=(N+1)/2$) will always remain motionless, as the forces acting on it from both sides cancel each other at all times. 
Futhermore, there is one configuration, where all forces are directed towards the center of the chain and one, in which the accelerations are directed away from it.  
According to (\ref{F_chain}) this is the case, when the initial electronic wavefunction has no nodes at all (i.e., the state $k=1$ at the lower edge of the band) and, when the sign of $\ket{\adBasis_k}$ changes from site to site (i.e. the state $k=N$ with the highest energy).
Generally, all forces reverse direction when changing the initial state from state $k$ to $(N+1)-k$.
A pair of atoms experiences an attractive force, if the wavefunction on the corresponding sites has equal sign and a repulsive force otherwise. The total acceleration of a particle due to its two neighbors is then determined by the relative magnitude of the wavefunction at neighboring sites (c.f. Eq. \ref{F_general}).

\begin{figure}
\centering
\includegraphics[width=0.75\textwidth]{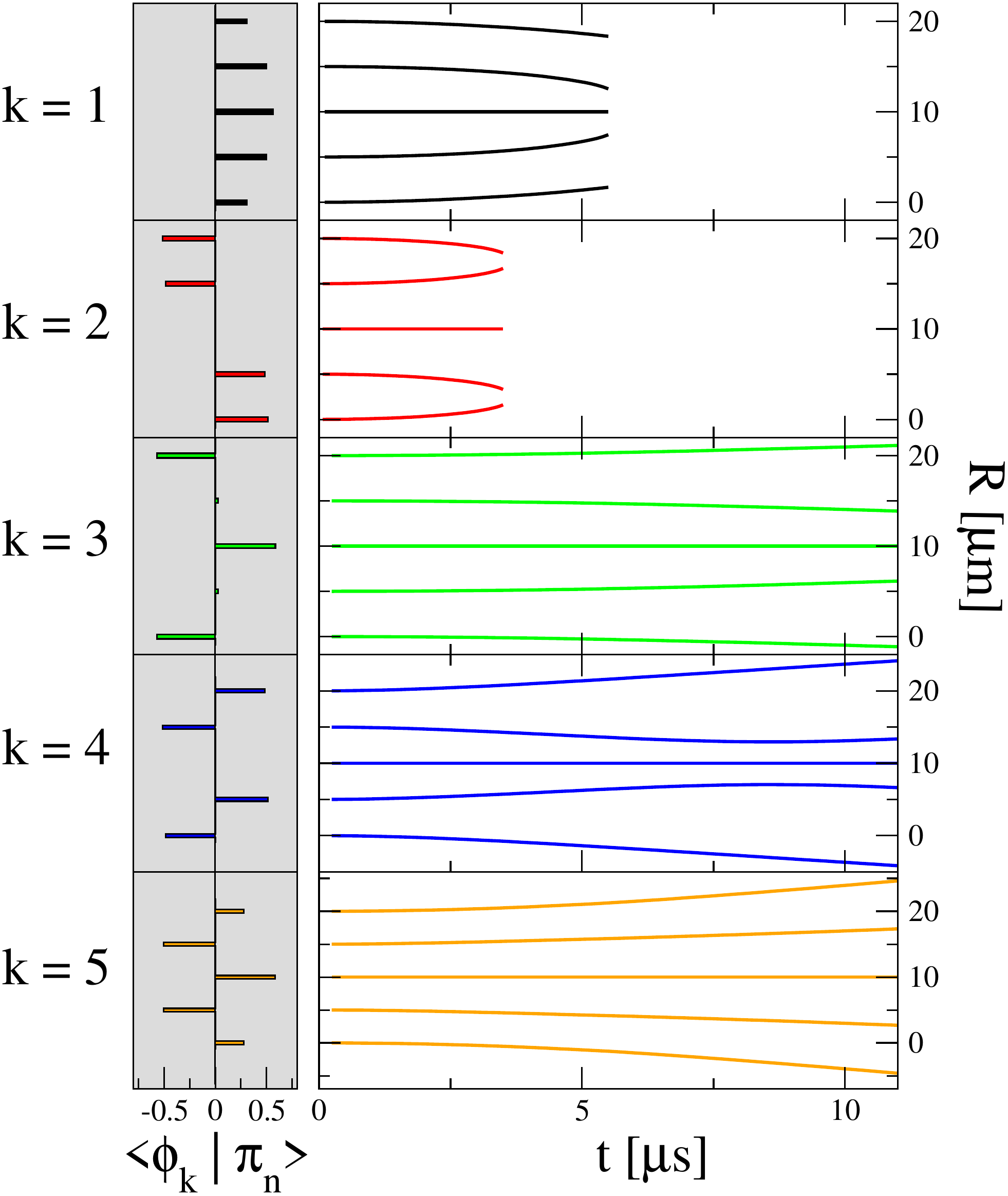}
\caption{
Left column: Electronic eigenstates of a regular chain with $N=5$ atoms with increasing energies from top to bottom. Right column: Trajectories of the particles, when the system is initially prepared in the respective eigenstate on the left column. For details, see text.
} \label{fig:5mer}
\end{figure}

For a chain with length $N=5$, Fig. \ref{fig:5mer} shows the numerically calculated trajectories, for the system prepared in the five possible electronic eigenstates $\adEnergy_k$, where now the interactions between all atoms have been taken into account. The graphs in the shaded areas show the corresponding eigenfunctions at the lattice sites $\bra{\adBasis_k}  \pi_n\rangle$, where the site index $n$ increases from bottom to top in each graph.

The trajectories starting from the two energetically lowest eigenstates show relatively fast collisions eventually leading to ionization of the Rydberg atoms, whereas for the two highest eigenstates no collisions occur, although for $k=4$ the trajectories of the three central atoms come close to each other. For $k=3$ Eq.\ref{F_chain} predicts that no forces at all are present in the system, which seems to contradict the numerical result. However, the analytical estimate for the forces was derived under the assumption of nearest neighbor interactions only. In the general case of the ''full'' $R^{-3}$ interaction, the excitonic band shows a marked asymmetry ($\adEnergy_1=-2.403\, V(\Abstand)$ and $\adEnergy_N = 1.803\, V(\Abstand)$ for $N \to \infty$) and the phase of the eigenfunctions is shifted due to the additional interactions from more distant sites. Thus, the $\bra{\adBasis_k}  \pi_n\rangle$ are nonzero at the sites next to the central atom, which in turn leads to a small but finite force between the particles.
%------------------------------------------------------------------------------------------------------------------------------------------
%------------------------------------------------------------------------------------------------------------------------------------------

\begin{figure}
\centering
\includegraphics[width=0.48\textwidth]{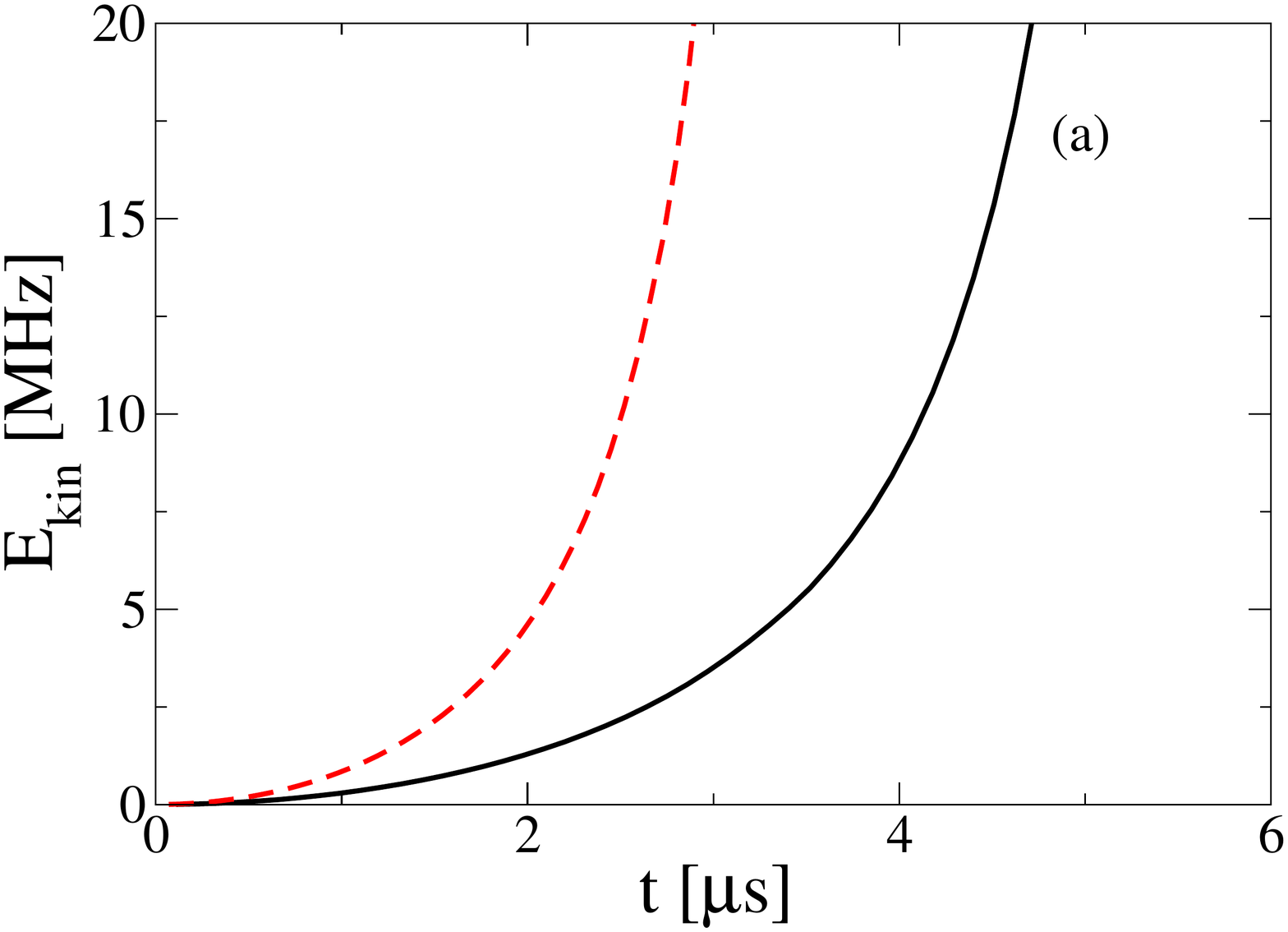} \hfill
\includegraphics[width=0.48\textwidth]{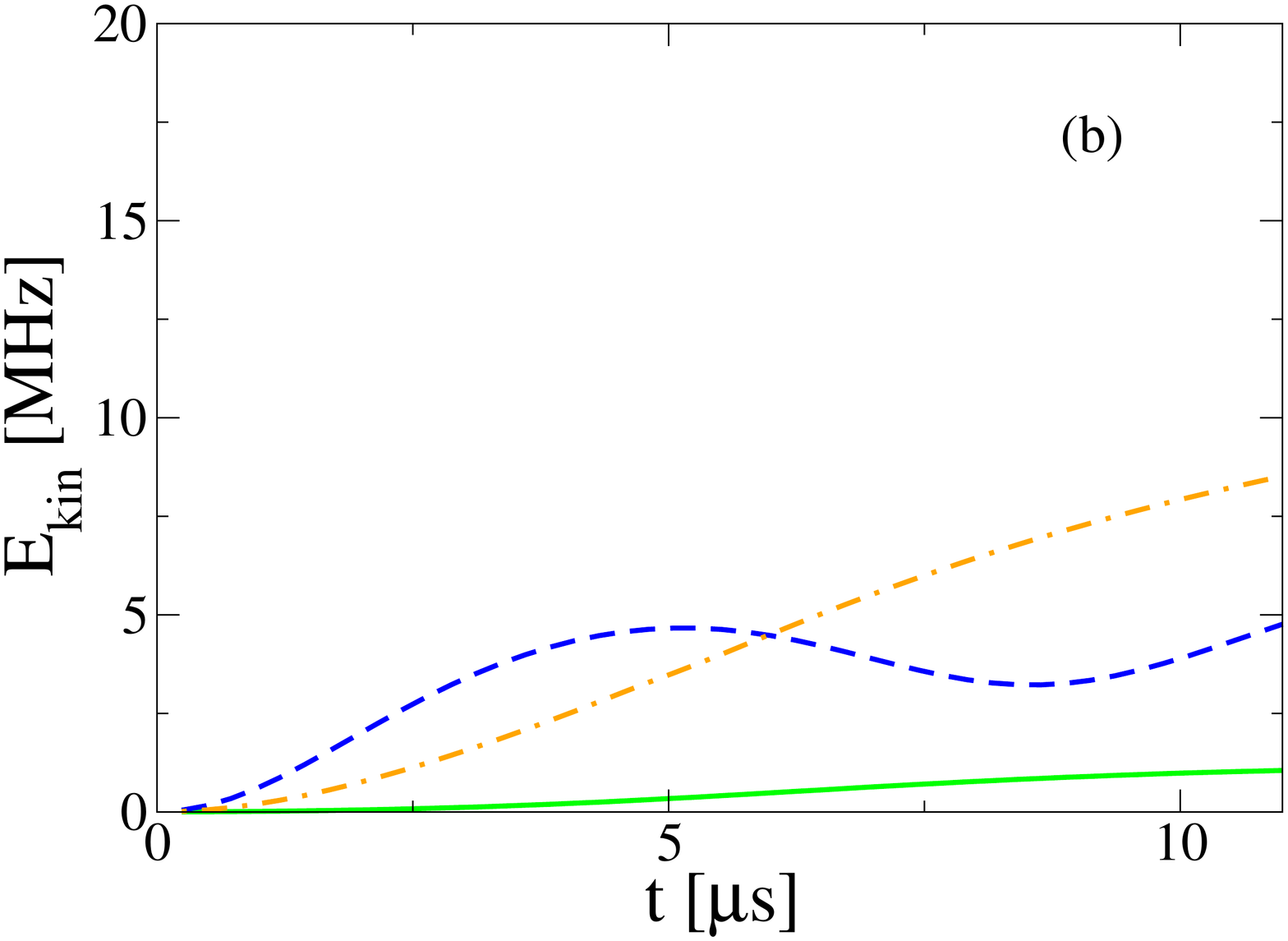}
\caption{
Time evolution of the kinetic energy for a regular chain of $N=5$ atoms, when the electronic state of the system is prepared in the five eigenstates. (a) Initial states $k=1$ (solid) and $k=2$ (dashed) that lead to fast collisions. (b) Initial states $k=3$ (solid), $k=4$ (dashed) and $k=5$ (dash-dotted) that show a slow dynamics.
}\label{fig:ek5}
\end{figure}

The dependence of the initial nuclear motion on the initial electronic state can also be seen in the time evolution of the kinetic energy of the system. 
For short times one can write $\Ekin(t) = (M/2)\sum v_n^2(t)$ with
 $v_n(t)\approx t \Kraft_n/M$.
Inserting the initial forces \eref{F_chain} one finds
\begin{equation}
\Ekin(t) \approx \frac{36}{N+1} \left(\frac{\mu^2}{\Abstand^4}\right)^2
\sin^2(\frac{\pi k}{N+1})\cdot \frac{t^2}{M} \, ,
\end{equation}
and $\Ekin(t) \approx 0$ for  $k = (N+1)/2$, if $N$ is odd.
We found good agreement of this formula with the numerical calculations shown in Fig. \ref{fig:ek5} for times $t< 1\,\mu$s. 
The fast collsions in the trajectories for the two lowest initial electronic states lead to a rapid increase of the kinetic energy (Fig. \ref{fig:ek5}(a)). This energy will eventually be available for ionizing the Rydberg atoms, when they approach each other.  
In contrast,  $\Ekin$ increases much slower for the three eigenstates that do not show rapidly colliding trajectories (Fig. \ref{fig:ek5}(b)). The slow ''collision'' that is seen in the trajectories for $k=4$ manifests itself in a non-monotonic time evolution of $\Ekin$ with a minimum at the time, when the central atoms reach their minimal distance. 
%------------------------------------------------------------------------------------------------------------------------------------------
%------------------------------------------------------------------------------------------------------------------------------------------

%%%%%%%%%%%%%%%%%%%%%%%%%%%%%%%%%%%%%%%%%%%%%%%%%%%%%%%%%%%%%%%%%%%%%%%%%%%%%%%%%%%%%%%%%%%%%%%%%%%%%%
\section{Off-diagonal disorder - chain with ''trap'' at one end}
\label{disorder}
In this section we will study the effect, when the regularity of the chain is broken. Since, the regularity of the mutual atomic interactions $V_{nm}$ (for $n\ne m$) is destroyed, this is an example of a system displaying off-diagonal disorder.
We will focus on the specific case, where all atoms, but the last two have a spacing $\Abstand$ and the distance between the last two particles (which we will call the ''trap atoms'') is $\altAbstand$.

In this case, there are generally no analytical solutions for the electronic eigenenergies and eigenstates, even if the interaction is restricted to nearest neighbor sites. However, if the eigenstates are determined numerically, Eq. \ref{F_general} can still be used to estimate the forces acting initially on the particles.

\begin{figure}
\centering
\includegraphics[width=0.75\textwidth]{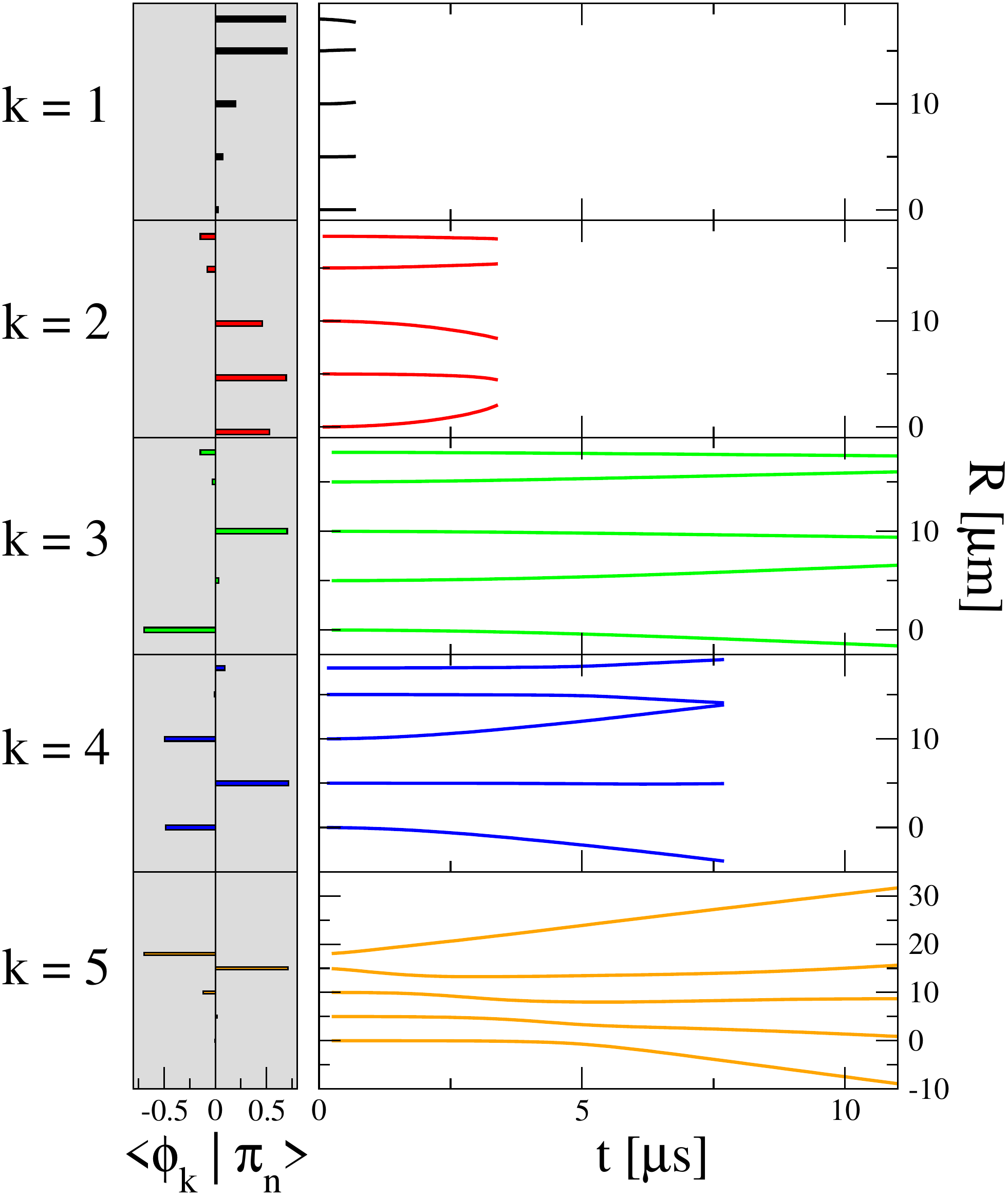}
\caption{
Same as Fig. \ref{fig:5mer}, but for a chain with trap at the upper end. The ratio of the distance of the trap atoms to the nearest neighbor distance in the regular part of the chain is $3/5$.
}
\label{fig:trap}
\end{figure}

For a chain of $N=5$ atoms, Fig. \ref{fig:trap} shows the trajectories together with the initial electronic eigenstates $\bra{\adBasis_k}  \pi_n\rangle$, where we have initially set $\altAbstand/\Abstand=3/5$. In this case, the dipole-dipole interaction experienced by the trap atoms is approximately 5 times larger than by the particles in the rest of the chain.

\begin{figure}
\centering
\includegraphics[width=0.48\textwidth]{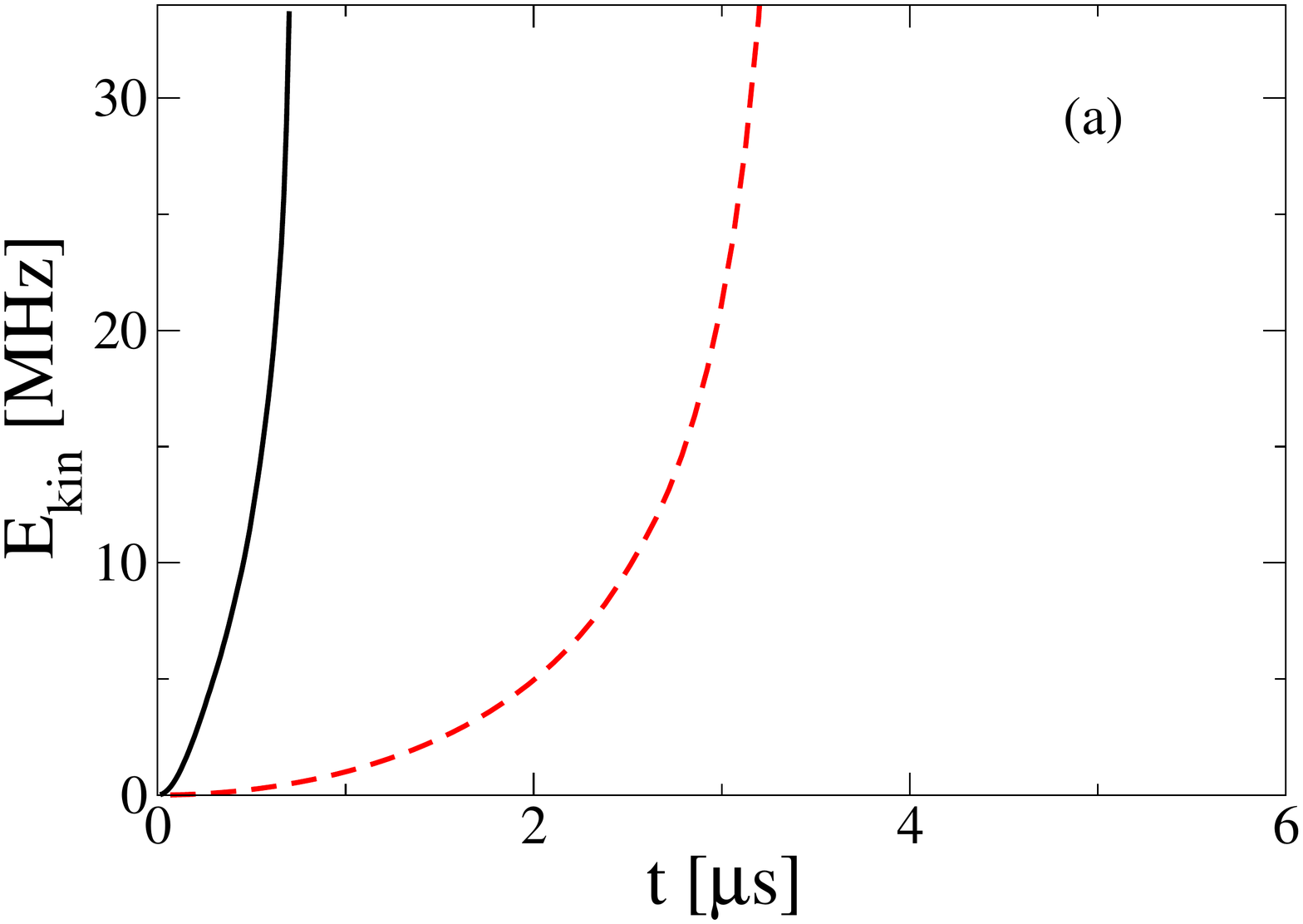} \hfill
\includegraphics[width=0.48\textwidth]{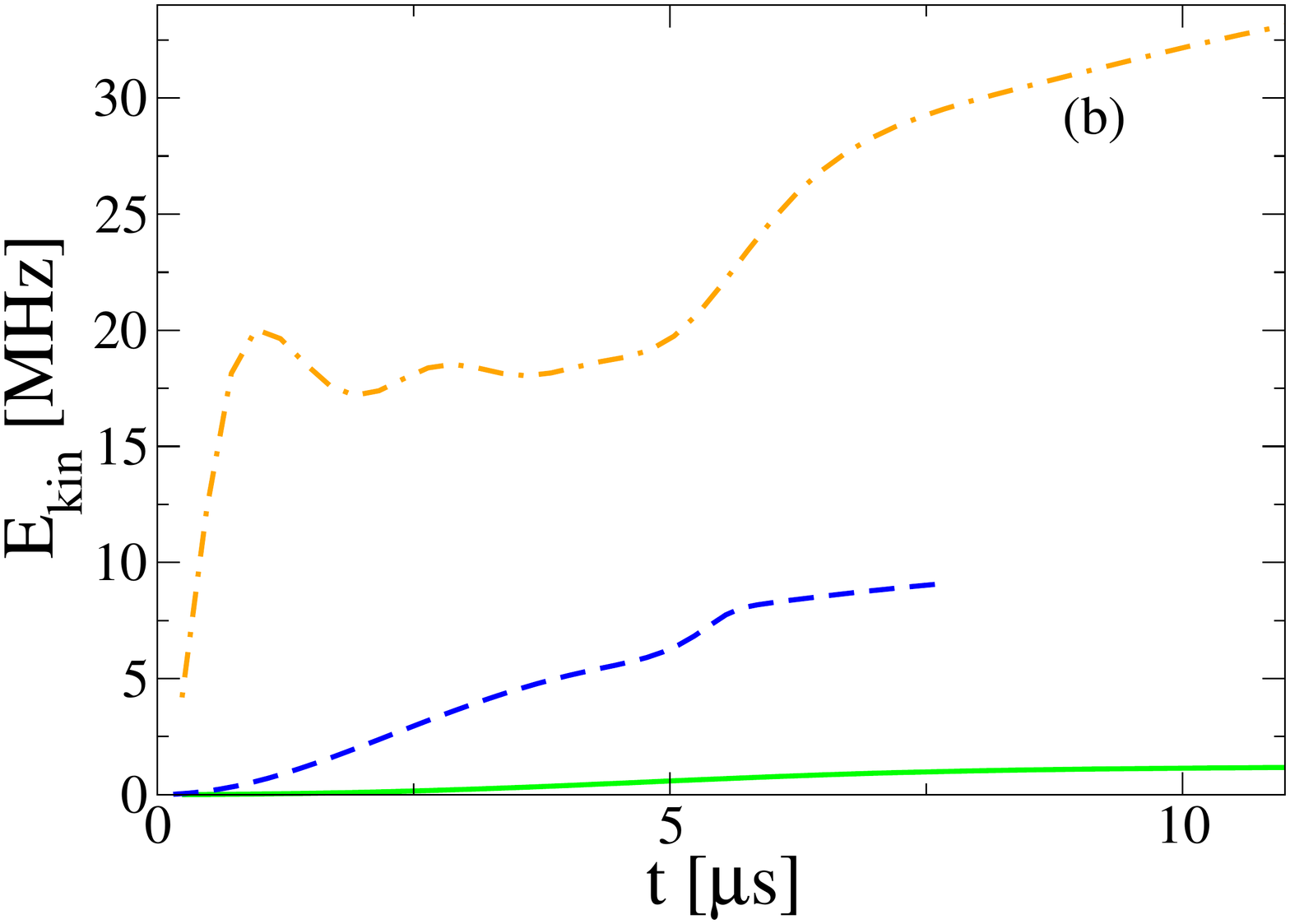}
\caption{
Same as Fig. \ref{fig:ek5}, but for a chain with a trap at one end. The ratio of the distance of the trap atoms to the nearest neighbor distance in the regular part of the chain is $3/5$.
}
\label{fig:ektrap}
\end{figure}

The eigenstate at the band edges ($k=1$ and $k=5$) are now strongly localized on the
trap atoms and can approximately be written as $(\ket{\pi_4} \pm \ket{\pi_5})/\sqrt{2}$, resembling  an attractive and a repulsive dimer state, respectively.
Consequently, the trajectories, when starting in these states, show, respectively,  a strong attraction and a strong repulsion of the trap atoms.  In the former case, the fast collsion is much better seen in the time evolution of the kinetic energy (Fig. \ref{fig:ektrap}(a)).

%------------------------------------------------------------------------------------------------------------------------------------------
%------------------------------------------------------------------------------------------------------------------------------------------
%------------------------------------------------------------------------------------------------------------------------------------------
\begin{figure}
\centering
\includegraphics[width=0.9\textwidth]{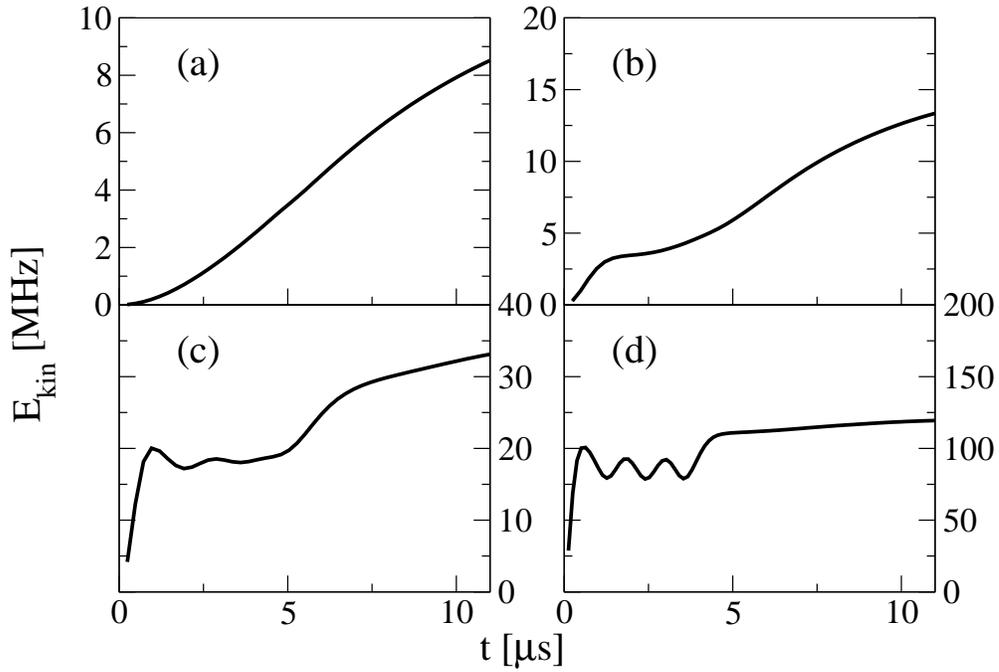}
\caption{
Time evolution of the kinetic energy for a chain of $N=5$ atoms with trap at one end starting from the $k=5$ electronic eiegnstate for different ratios $\gamma=y/x$ of the distance of the trap atoms to the nearest neighbor distance in the regular part of the chain. (a) $\gamma=1$ (regular chain), (b) $\gamma=4/5$, (c) $\gamma=3/5$ and (d) $\gamma=2/5$.
}
\label{fig:disorder}
\end{figure}

%------------------------------------------------------------------------------------------------------------------------------------------
Interestingly, if the system is prepared in the repulsive dimer state ($k=5$), the ''dimer character'' is transferd from pair to pair. When two atoms come close to each other, they constitute a new repulsive dimer, such that the initial momentum of the first dimer is transfered through the chain. This effect can be enhanced, if the distance of the trap atoms is reduced further. Fig. \ref{fig:disorder} shows the time evolution of the kinetic energy for differnt ratios $\altAbstand/\Abstand$ ranging from 1 (i.e., completely regular chain) to $2/5$, where the electronic eigenstate is partically completely localized on the trap. First, the kinetic energy of the system increases with decreasing distance of the trap atoms, as the initial electronic (potential) energy is dominated by the distance of the trap atoms. Second,  $\Ekin$ shows an increasingly non-monotonic behavior leading to pronounced  oscillations for small $\altAbstand/\Abstand$, where the minima correspond to times, when the distance of two atoms becomes minimal, i.e., a new dimer is formed. 

In the cases, when the system is prepared in on of the eigenstates corresponding to energies in the center of the exciton band, the excitation is delocalized over the regular part of the chain.
The trajectories (Fig. \ref{fig:trap}) and corresponding kinetic energies (Fig. \ref{fig:ektrap}(a), (b)) show a similar behavior like for a completely regular chain, with the exception that in the slow collision process for $k=4$ the Rydberg atoms approach each other so closely, that the assumption of non overlapping wavefunctions breaks down, so that this process has to be treated completely quantum mechanically.

%------------------------------------------------------------------------------------------------------------------------------------------
%------------------------------------------------------------------------------------------------------------------------------------------

\section{Conclusions}
\label{concl}
We have investigated the nuclear motion in a chain of Rydberg atoms, which is induced by resonant dipole-dipole interactions that trigger the energy transfer between two energetically close Rydberg states. 

The motion of the atoms depends crucially on the electronic state, in which the system is prepared. We have considered the particular case, when the system is prepaered in eigenstates of the exciton band. For a completely regular chain, we have given simple estimates for the initial forces and the time evolution of the kinetic energy for short times. 
In particular, we have shown that the magnitude of the forces acting initially is inversely proportional to the number of atoms, due to delocalization of the electronic excitation over the whole chain.
This delocalization of energy is also responsible for phenomena like exchange narrowing \cite{WaEiBr__} or superradiance found in molecular aggregates.

Furthermore, we have studied the influence of so called ''off-diagonal disorder'' in a simple example, in which we have distrubed the regularity of the system by placing two atoms closer together than the particles in the rest of the chain. In this case, some electronic eigenfunctions are strongly localized on the perturbed sites and the degree of localization increases with increasing irregularity.  In our example, where the irregularity leads to larger initial potential energy a much faster nuclear motion results in the system and a transfer of the initial momentum of the atoms forming dimer through the chain. 

\ack{We thank U. Saalmann for bringing Ref. \cite{Tu90_1061_} to our attention.}

%\input{appendix.tex}

%%%%%%%%%%%%%%%%%%%%%%%%%%%%%%%%%%%%%%%%%%%%%%%%%%%%%%%%%%%%%%%%%%%%%%%%%%%%%%%%%%%%%%%%/home/eisfeld/tex/PaperBib_alt/PaperBibPPP,
\bibliographystyle{unsrt}
%\bibliography{PaperBib,Rydberg-Dresden}% Produces the bibliography via BibTeX.

\begin{thebibliography}{10}

\bibitem{AmVaGr00__}
H.~van Amerongen, L.~Valkunas, and R.~van Grondelle,
\newblock {\em {Photosynthetic Excitons}}.
\newblock World Scientific, Singapore (2000)

\bibitem{Ko96__}
T.~Kobayashi, editor,
\newblock {\em {J-Aggregates}}.
\newblock World Scientific, Singapore (1996)

\bibitem{WeSt03_125201_}
M.~Wewer and F.~Stienkemeier,
%\newblock {Molecular versus Excitonic Transitions in {PTCDA} Dimers and
%  Oligomers Studied by Helium Nanodroplet Isolation Spectroscopy}.
\newblock {\em Phys. Rev. B}, {\bf 67},125201 (2003).

\bibitem{Da62__}
A.S. Davydov,
\newblock {\em {Theory of Molecular Excitons}},
\newblock McGraw-Hill (1962)

\bibitem{FrTe38_861_}
J.~Franck and E.~Teller,
%\newblock {Migration and Photochemical Action of Excitation Energy in
 % Crystals}.
\newblock {\em J. Chem. Phys.},{\bf 6} 86 (1938)

\bibitem{Fr31_1276_}
J.~Frenkel,
%\newblock {On the Transformation of Light into Heat in Solids. {II}}.
\newblock {\em Phys. Rev.}, {\bf 37}, 1276 (1931)

\bibitem{EiBr06_113003_}
A.~Eisfeld and J.~S. Briggs,
%\newblock {Absorption Spectra of Quantum Aggregates Interacting via Long-Range
%  Forces}.
\newblock {\em Phys. Rev. Lett.}, {\bf 96}, 113003  (2006)

\bibitem{MoCoTo98_253_}
I.~Mourachko, D.~Comparat, F.~de~Tomasi, A.~Fioretti, P.~Nosbaum, V.~M. Akulin,
  and P.~Pillet,
%\newblock {Many-Body Effects in a Frozen Rydberg Gas}.
\newblock {\em Phys. Rev. Lett.}, {\bf 80}, 253 (1998)

\bibitem{AnVeGa98_249_}
W.~R. Anderson, J.~R. Veale, and T.~F. Gallagher,
%\newblock Resonant dipole-dipole energy transfer in a nearly frozen rydberg
 % gas.
\newblock {\em Phys. Rev. Lett.}, {\bf 80}, 249 (1998)

\bibitem{AfBoVa04_233001_}
K.~Afrousheh, P.~Bohlouli-Zanjani, D.~Vagale, A.~Mugford, M.~Fedorov, and
  J.~D.~D. Martin,
%\newblock Spectroscopic observation of resonant electric dipole-dipole
%  interactions between cold rydberg atoms.
\newblock {\em Phys. Rev. Lett.}, {\bf 93},  233001 (2004)

\bibitem{WeAmOl06_37_}
S.~Westermann, T.~Amthor, A.~L. de~Oliveira, J.~Deiglmayr, M.~Reetz-Lamour, and
  M.~Weidem\"uller,
%\newblock Dynamics of resonant energy transfer in a cold rydberg gas.
\newblock {\em European Physical Journal D}, {\bf 40}, 37 (2006)

\bibitem{RoHeTo04_042703_}
F.~Robicheaux, J.~V. Hern\'{a}ndez, T.~Top\c{c}u, and L.~D. Noordam,
%\newblock Simulation of coherent interactions between rydberg atoms.
\newblock {\em Physi. Rev. A}, {\bf 70}, 042703  (2004)

\bibitem{MueBlAm07_090601_}
O.~M\"ulken, A.~Blumen, T.~Amthor, C.~Giese, M.~Reetz-Lamour, and M.~Weidem\"uller,
%\newblock {Survival probabilities in coherent exciton transfer with trapping}.
\newblock {\em Phys. Rev. Lett.}, {\bf 99}, 090601 (2007)

\bibitem{AmReWe07_023004_}
T.~Amthor, M.~Reetz-Lamour, S.~Westermann, J.~Denskat, and M.~Weidem\"uller,
%\newblock Mechanical effect of van der waals interactions observed in real time
 % in an ultracold rydberg gas.
\newblock {\em Phys. Rev. Lett.}, {\bf 98}, 023004 (2007)

\bibitem{AmReGi07_054702_}
T.~Amthor, M.~Reetz-Lamour, C.~Giese, and M.~Weidem\"uller,
%\newblock {Modeling many-particle mechanical effects of an interacting Rydberg
 % gas}.
\newblock {\em Phys. Rev. A}, {\bf 76}, 054702 (2007)

\bibitem{LiTaGa05_173001_}
W.~H. Li, P.~J. Tanner, and T.~F. Gallagher,
%\newblock Dipole-dipole excitation and ionization in an ultracold gas of
%  rydberg atoms.
\newblock {\em Phys. Rev. Lett.}, {\bf 94}, 173001 (2005)

\bibitem{LiTaJa06_27_}
W.~Li, P.~J. Tanner, Y.~Jamil, and T.~F. Gallagher,
%\newblock Ionization and plasma formation in high n cold rydberg samples.
\newblock {\em European Physical Journal D}, {\bf 40}, 27 (2006)

\bibitem{FiCoDr99_1839_}
A.~Fioretti, D.~Comparat, C.~Drag, T.~F. Gallagher, and P.~Pillet,
%\newblock Long-range forces between cold atoms.
\newblock {\em Phys. Rev. Lett.}, {\bf 82}, 1839 (1999)

\bibitem{WaRoJo07_3693_}
M.~L. Wall, F.~Robicheaux, and R.~R. Jones,
%\newblock {Controlling atom motion through the dipole-dipole force}.
\newblock {\em J. Phys. B}, {\bf 40}, 3693 ( 2007)

\bibitem{Tu90_1061_}
J.~C. Tully.
%\newblock {Molecular-Dynamics With Electronic-Transitions}.
\newblock {\em J. Chem. Phys.}, {\bf 93}, 1061, (1990)

\bibitem{WaEiBr__}
P.B. Walczak, A.~Eisfeld, and J.~S. Briggs,
%\newblock Exchange narrowing of the {J}-band of molecular dye aggregates.
\newblock {\em J. Chem. Phys.}, accepted (2007)

\end{thebibliography}

\end{document}